\documentclass[a4paper]{article}
\usepackage{amsmath, epsfig, tabularx}
\newcommand{\cali}[1]{\ensuremath{\mathcal{#1}}}
\newcommand{\pdif}[2]{\ensuremath{\frac{\partial #1}{\partial #2}}}
\newcommand{\pdiftwo}[2]{\ensuremath{\frac{\partial^2 #1}{\partial
#2^2}}} 

\title{Hedging in Field Theory Models of the Term Structure}
\author{Belal E. Baaquie and Marakani Srikant\\Department of Physics,
  National University of Singapore\\phybeb@nus.edu.sg,
  srikant@srikant.org} 
\begin{document}
\maketitle
\begin{abstract}
  We use path integrals to calculate hedge parameters and efficacy of
  hedging in a quantum field theory generalization of the Heath,
  Jarrow and Morton (HJM) term structure model which parsimoniously
  describes the evolution of imperfectly correlated forward rates. We
  also calculate, within the model specification, the effectiveness of
  hedging over finite periods of time. We use empirical estimates for
  the parameters of the model to show that a low dimensional hedge
  portfolio is quite effective. 
\end{abstract}
\section{Introduction}
The first interest models were spot rate models which had only one
factor which implied that the prices of all bonds\footnote{In this
  paper, we only use zero coupon bonds, hence all references to bonds
  are to zero coupon bonds} were perfectly correlated. This was
observed not to be the case in practice. This observation led Heath,
Jarrow and Morton \cite{HJM} to develop their famous model (heceforth
called the HJM model) where the forward rate curve was influenced by
more than one factor. This enabled bond prices to have imperfect
correlation. However, for an $N$ factor HJM model, this still meant
that the movements in the price of $N$ bonds would determine the
movements in the prices of all other bonds. This would enable one to
hedge any instrument with $N$ bonds within the framework of this
model. However, this does not again seem to be the case in practice.
In fact, if taken to be exact, a two factor HJM model would seem to
imply that one can hedge a thirty year treasury bond with three month
and six month bills which is not reasonable. Hence, there has been
much interest in developing models which do not have this problem. One
possibility is to use an infinite factor HJM model as pointed out in
Cohen and Jarrow \cite{Cohen} but it is well known that estimating the
parameters of even a two or three factor HJM model from market data is
very difficult. In contrast, the estimation of parameters for
different field theory models has been discussed in Baaquie and
Srikant \cite{BaaqSri} and is seen to be more effective.

These observations led Kennedy \cite{Ken}, Santa-Clara and
Sornette \cite{Pedro} and Goldstein \cite{Gold} to come up with random
field models which allowed imperfect correlations across all the
bonds. Baaquie \cite{Baaquie} furthered this development by putting
all these models into a field theory framework which allows for
the use of a large body of theoretical and computational methods
developed in physics to be applied to this problem. 

\section{A Brief Summary of the HJM and Field Theory Models}
In the HJM model the forward rates are given by 
\begin{equation}
  \label{eq:hjmdef}
  f(t, x) = f(t_0, x) + \int_{t_0}^t dt' \alpha(t', x) + \sum_{i=1}^K \int_{t_0}^t
  dt' \sigma_i(t',x) dW_i(t')
\end{equation}
where $W_i$ are independent Wiener processes. We can also write this
as 
\begin{equation}
  \label{eq:hjmdefint}
  \pdif{f(t,x)}{t} = \alpha(t,x) + \sum_{i=1}^K \sigma_i(t,x) \eta_i(t)
\end{equation}
where $\eta_i$ represent independent white noises. The action
functional, is 
\begin{equation}
  \label{eq:hjmaction}
  S[W] = -\frac{1}{2}  \sum_{i=1}^K \int dt \eta_i^2(t)
\end{equation}
We can use this action to calculate the generating functional which is
\begin{eqnarray}
\label{eq:z}
Z[j,t_1,t_2]&=&\int {\cal D}W e^{\sum_{i=1}^K\int_{t_1}^{t_2}dt
    j_i(t)W_i(t)}e^{S_0[W,t_1,t_2]}\nonumber  \\
    &=&e^{\frac{1}{2}\sum_{i=1}^K \int_{t_1}^{t_2}dt j_i^2(t)}
\end{eqnarray}

We now review Baaquie's field theory model presented in \cite{Baaquie}
with constant rigidity. Baaquie proposed that the forward rates being
driven by white noise processes in (\ref{eq:hjmdefint}) be replaced by
considering the forward rates itself to be a quantum field. To
simplify notation, we write the evolution equation in terms of the
velocity quantum field $A(t, x)$, and which yields
\begin{equation}
  \label{eq:baaquiedef}
  f(t, x) = f(t_0, x) + \int_{t_0}^t dt' \alpha(t', x) + \sum_{i=1}^K \int_{t_0}^t
  dt' \sigma_i(t',x) A_i(t',x)
\end{equation}
or 
\begin{equation}
  \label{eq:baaquiedefdif}
  \pdif{f(t,x)}{t} = \alpha (t,x) + \sum_{i=1}^K \sigma_i(t,x) A_i(t,x)
\end{equation}
The main extension to HJM is that $A$ depends on $x$ as well as $t$
unlike $W$ which only depends on $t$. 

While we can put in many fields $A_i$, it was shown in Baaquie and
Srikant \cite{BaaqSri} that the extra generality brought into the
process due to the extra argument $x$ makes one field sufficient.
Hence, in future, we will drop the subscript for $A$.

Baaquie further proposed that the field $A$ has the free (Gaussian)
free  field action functional \cite{Baaquie}
\begin{equation}
  \label{eq:adef}
  S = -\frac{1}{2} \int_{t_0}^{\infty} dt \int_t^{t+T_{FR}} dx \left(A^2 +
    \frac{1}{\mu^2} \left(\pdif{A}{x}\right)^2\right)
\end{equation}
with Neumann boundary conditions imposed at $x=t$ and
$x=t+T_{FR}$. This makes the action equivalent (after an integration
by parts where the surface term vanishes) to 
\begin{equation}
  \label{eq:adef1}
  S = -\frac{1}{2} \int_{t_0}^\infty dt \int_t^{t+T_{FR}} dx A(t, x)\left(1
    - \frac{1}{\mu^2}\pdiftwo{}{x}\right)A(t, x) 
\end{equation}
This action has the partition function 
\begin{equation}
  \label{eq:partition}
  Z[j] = \exp\left(\int_0^{t_1} dt \int_t^{t+T_{FR}} dx dx' j(t,x)
    D(x-t,x'-t) j(t,x')\right)
\end{equation}
with 
\begin{equation}
  \label{eq:D}
  \begin{split}
    D(\theta,\theta';T_{FR})&=\mu\frac{\cosh\mu\left(T_{FR}-|\theta-\theta'|\right) + \cosh\mu\left(T_{FR}-(\theta+\theta')\right)} {2\sinh\mu T_{FR}}\\   
    &= D(\theta',\theta;T_{FR})\quad : \quad
    \mathrm{Symmetric~Function~of}~\theta,\theta'
\end{split}
\end{equation}
where $\theta = x-t$ and $\theta' = x'-t$. We can calculate
expectations and correlations using this partition function.  Note
that due to the boundary conditions imposed, the inverse of the
differential operator $D$ actually depends only the difference $x-t$.
The above action represents a Gaussian random field with covariance
structure $D$. In \cite{Baaquie}, a different form was found as the
boundary conditions used were Dirichlet with the endpoints integrated
over. This boundary condition is in fact equivalent to the Neumann
condition which leads to the much simpler propagator above. In the
limit $T_{FR} \rightarrow \infty$ which we will usually take, the
propagator takes the simple form $\mu e^{-\mu \theta_>} \cosh \mu
\theta_<$ where $\theta_>$ and $\theta_<$ stand for $\max(\theta,
\theta')$ and $\min(\theta, \theta')$ respectively.

When $\mu \rightarrow 0$, this model should go over to the HJM model.
This is indeed seen to be the case as it is seen that $\lim_{\mu
  \rightarrow 0} D(\theta, \theta'; T_{FR}) = \frac{1}{T_{FR}}$. The
extra factor of $T_{FR}$ is irrelevant as it is due to the freedom we
have in scaling $\sigma$ and $D$. The $\sigma$ we use for the
different models are only comparable after $D$ is
normalized\footnote{This freedom exists since we can always make the
  transformation $\sigma(\theta) \sim \eta(\theta) \sigma(\theta)$ and
  $D(\theta, \theta') \sim D(\theta, \theta')/(\eta(\theta)
  \eta(\theta'))$ without affecting any result}. On normalization, the
propagator for both the HJM model and field theory model in the limit
$\mu \rightarrow 0$ is one showing that the two models are equivalent
in this limit.

\section{Hedging}
The main aim of hedging is to reduce one's exposure to risk. There are
many ways to define risk \cite{Risks}. For bonds, the main risks are
changes in interest rates and the risk of default. In this paper, we
are only dealing with default-free bonds so that the only source of
risk is the change in interest rates. 

For the purposes of this paper, we define risk to be the standard
deviation or variance of final value. Hence, when we hedge a certain
instrument, we are trying to create a portfolio of the hedged and
hedging instruments which minimizes the overall variance of the
portfolio. In the case of a N-factor HJM model, perfect hedging (i.e.,
a zero portfolio variance) is achievable once any N independent
hedging instruments are used. However, the difficulties introduced by
the infinite number of factors in the field theory models has resulted
in their being very little literature on this important subject, a
notable exception being the measure valued trading strategy developed
in Bj\"ork, Kabanov and Runggaldier\cite{BKR}.

In the fourth section of this paper, we will consider instantaneous
hedging which is important for theoretical purposes. We will calculate
the maximum reduction in variance for a finite number of hedging
instruments and the hedge ratios (the amount of hedging instrument
that requires to be used) that result in this maximization.  This will
show us how well the model can be approximated by a finite number of
factors.  We will then use the constant rigidity model fitted with
empirical data to estimate the reduction in the variance of an
optimally hedged portfolio as the number of hedging instruments are
increased.  We will see that a relatively small number of hedging
instruments gives good results.  We will also show that the results
reduced to well known textbook ones as in Jarrow and
Turnbull\cite{JTtext} when we go to the degenerate case of one-factor
HJM model where all the forward rate innovations are perfectly
correlated. We will also perform the same calculations using the
propagator estimated from empirical data.

In the third section, we will consider finite time hedging which is
important in practice. This is because continuous hedging cannot be
done in practice due to the presence of transaction costs. We will see
how the hedging performance found in the second section changes
as the time between rebalancings is increased. The entire analysis
here is to investigate how portfolios of bonds behave in such models. 

\section{Instantaneous Hedging}
In instantaneous hedging, we are considering a hedging portfolio which
is rebalanced continuously in time. Hence, we are only considered with
the instantaneous variance of the portfolio. This can be calculated
for an arbitrary portfolio by using the fact that the covariance of
the innovations in the forward rates is given by 
\begin{equation}
  \label{eq:covariance}
  \sigma(\theta) D(\theta, \theta') \sigma(\theta')
\end{equation}
in the field theory model. We will only present the hedging of zero
coupon bonds in this section though it will be seen that the results
can be easily extended to other instruments. In the first subsection,
we will present the theoretical derivation of the hedge ratios and
reduced variance for the hedging of a zero coupon bond with other zero
coupon bonds. In the second subsection, we use the empirically fitted
$\sigma$ and $D(\mu)$ (from (\ref{eq:D})) discussed in Baaquie and
Srikant \cite{BaaqSri} for the constant rigidity action as well as the
non-parametric estimate for $\sigma$ and $D$ which is directly
obtained from the market correlation matrix of the innovations in
forward rates to calculate the semi-empirical reduction in variance.
In the third and fourth subsections, we will carry out similar
calculations when hedging zero coupon bonds with futures on zero
coupon bonds. This is much more realistic in practice as hedging with
futures is relatively cheap.

\subsection{Hedging bonds with other bonds} \label{theo_hedging}
We now consider the hedging of one bond maturing at $T$ with $N$ other
bonds maturing at $T_i, 1\le i \le N$. If one of the $T_i = T$, then
the solution is trivial since it is the same bond. The hedge is then
just to short the same bond giving us a zero portfolio with obviously
zero variance. Since this solution is uninteresting, we assume that
$T_i \ne T\,\, \forall i$. The hedged portfolio $\Pi(t)$ can then be
represented as 
\begin{equation}
  \Pi(t) = P(t,T) + \sum_{i=1}^N \Delta_i P(t,T_i) \nonumber
\end{equation}
where $\Delta_i$ denotes the amount of the $i^{th}$ bond $P(t,T_i)$
included in the hedged portfolio. Note the value of bonds $P(t,T)$ and
$P(t,T_i)$ are determined by observing their market values at time
$t$. It is the instantaneous {\it change} in the portfolio value that
is stochastic. Therefore, the variance of this change is computed to
ascertain the efficacy of the hedge portfolio.

We first consider the variance of the value of an individual bond in
the field theory model. The definition $P(t, T) = \exp{(-\int_t^T dx
  f(t, x))}$ for zero coupon bond prices implies that
\begin{eqnarray}
\frac{dP(t,T)}{P(t,T)} &=& f(t, t)dt - \int_0^{T-t} d\theta df(t, \theta) \nonumber
\\ &=& \left( r(t)  - \int_0^{T-t} d\theta \alpha(\theta) -
  \int_0^{T-t} d\theta \sigma(\theta) A(t, \theta) \right) dt \nonumber  
\end{eqnarray}
and $E\left[ \frac{dP(t, T)}{P(t,t)} \right] = \left( r(t) - \int_0^{T-t}
  d\theta \alpha(\theta) \right) dt$ since $E[A(t, \theta)] = 0$. Therefore
\begin{equation}
\frac{dP(t, T)}{P(t,T)} - E\left[ \frac{dP(t, T)}{P(t,T)} \right] =  -dt\int_0^{T-t} d\theta \sigma(\theta) A(t, \theta) 
\end{equation}
Squaring this expression and invoking the result that $E[A(t, \theta) A(t,
\theta')] = \delta(0) D(\theta,\theta';T_{FR}) = \frac{D(\theta,\theta';T_{FR})}{dt}$
results in the instantaneous bond price variance
\begin{equation}
Var[dP(t,T)] = dt P^2(t, T) \int_0^{T-t} d\theta \int_0^{T-t} d\theta' \sigma(\theta) D(\theta,\theta';T_{FR}) \sigma(\theta')
\end{equation}  

As an intermediate step, the instantaneous variance of a bond
portfolio is considered. For a portfolio of bonds, $\hat{\Pi}(t) =
\sum_{i=1}^N \Delta_i P(t,T_i)$, the following results follow directly
\begin{equation}
d\hat{\Pi}(t) - E[d\hat{\Pi}(t)] = -dt \sum_{i=1}^N \Delta_i P(t,
T_i) \int_0^{T_i-t} d\theta \sigma(\theta) A(t, \theta)
\end{equation}  
and 
\begin{equation}
Var[d\hat{\Pi}(t)] = dt \sum_{i=1}^N
\sum_{j=1}^N \Delta_i \Delta_j P(t,T_i) P(t,T_j) \int_0^{T_i-t}
d\theta \int_0^{T_j-t} d\theta' \sigma(\theta)
D(\theta,\theta';T_{FR}) \sigma(\theta')  
\label{eq:bondvariance}
\end{equation}

The (residual) variance of the hedged portfolio 
\begin{equation}
\Pi(t) = P(t,T) + \sum_{i=1}^N \Delta_i P(t,T_i)
\end{equation}
may now be computed in a straightforward manner. For notational
simplicity, the bonds $P(t,T_i)$ (being used to hedge the original
bond) and $P(t,T)$ are denoted $P_i$ and $P$ respectively. Equation
(\ref{eq:bondvariance}) implies the hedged portfolio's variance equals
the final result shown below
\begin{equation}
  \begin{split}
    \label{eq:hedgingbonds}
    &P^2 \int_0^{T-t} d\theta \int_0^{T-t} d\theta' \sigma(\theta) \sigma
    (\theta') D(\theta,\theta';T_{FR}) \\
    +& 2 P \sum_{i=1}^N \Delta_i P_i \int_0^{T-t} d\theta \int_0^{T_i-t} d\theta'
    \sigma(\theta) \sigma (\theta') D(\theta,\theta';T_{FR})  \\
    +& \sum_{i=1}^N \sum_{j=1}^N \Delta_i \Delta_j P_i P_j
    \int_0^{T_i-t} d\theta     \int_0^{T_j-t} d\theta' \sigma(\theta)
    D(\theta,\theta';T_{FR}) \sigma(\theta')
  \end{split}
\end{equation}
Note that the residual variance depends on the correlation structure
of the innovation in forward rates described by the propagator $D$.
Ultimately, the effectiveness of the hedged portfolio is an empirical
question since perfect hedging is not possible without shorting the
original bond.  This empirical question is addressed in the next
subsection where the propagator calibrated to market data is used to
calculate the effectiveness. Minimizing the residual variance in
equation (\ref{eq:hedgingbonds}) with respect to the hedge parameters
$\Delta_i$ is an application of standard calculus. We introduce the
following notation for simplicity.
\begin{eqnarray}
  L_i &=& P P_i \int_0^{T-t} d\theta \int_0^{T_i-t} d\theta'
  \sigma(\theta) \sigma(\theta') D(\theta,\theta';T_{FR}) \nonumber \\ 
  M_{ij}  &=& P_i P_j \int_0^{T_i-t} d\theta \int_0^{T_j-t} d\theta'
  \sigma(\theta) \sigma(\theta') D(\theta,\theta';T_{FR})\ \nonumber 
\end{eqnarray}
$L_i$ is the covariance between the innovations in the hedged bond and
the $i$th hedging bond and $M_{ij}$ is the covariance between the
innovations of the $i$th and $j$th hedging bond.

The above definitions allow the residual variance in equation
(\ref{eq:hedgingbonds}) to be succinctly expressed as
\begin{eqnarray}
  \label{eq:res}
  &&P^2 \int_0^{T-t} d\theta \int_0^{T-t} d\theta' \sigma(\theta) \sigma(\theta') D(\theta,\theta';T_{FR}) 
  + 2  \sum_{i=1}^N \Delta_i L_i + \sum_{i=1}^N \sum_{j=1}^N \Delta_i \Delta_j M_{ij} 
\end{eqnarray}
The hedge parameters in the field theory model can now be evaluated
using basic calculus and linear algebra to obtain 
\begin{equation}
  \label{eq:resultdelta}
  \Delta_i = - \sum_{j=1}^N L_j M_{ij}^{-1} 
\end{equation}
and represent the optimal amounts of $P(t,T_i)$ to include in the
hedge portfolio when hedging $P(t,T)$.

Putting the result into (\ref{eq:hedgingbonds}), we see that the
variance of the hedged portfolio equals
\begin{eqnarray}
  \label{eq:result_var}
  V &=& P^2 \int_0^{T-t} d\theta \int_0^{T-t} d\theta' \sigma(\theta)
  \sigma(\theta') D(\theta,\theta';T_{FR}) - \sum_{i=1}^N \sum_{j=1}^N L_i M_{ij}^{-1} L_j 
\end{eqnarray}
which declines monotonically as $N$ increases.

The residual variance enables the effectiveness of the hedged
portfolio to be evaluated. Therefore, this result is the basis
for studying the impact of including different bonds in the hedged
portfolio as illustrated in the next subsection. For $N=1$, the
hedge parameter reduces to
\begin{eqnarray}
\Delta_1 &=& - \frac{P}{P_1} \left( \frac{ \int_0^{T-t} d\theta \int_0^{T_1-t} d\theta'
\sigma(\theta) \sigma(\theta') D(\theta,\theta';T_{FR})}{
 \int_0^{T_1-t} d\theta \int_0^{T_1-t} d\theta' \sigma(\theta) \sigma(\theta') D(\theta,\theta';T_{FR})} \right) \label{eq:hed} 
\end{eqnarray}
To obtain the HJM limit, we let the propagator equal one. The hedge
parameter in equation (\ref{eq:hed}) then reduces to
\begin{eqnarray}
\Delta_1  &=& - \frac{P}{P_1} \left( \frac{ \int_0^{T-t} d\theta \int_0^{T_1-t} d\theta'
\sigma(\theta) \sigma(\theta') }{ \left( \int_0^{T_1-t} d\theta
  \sigma(\theta) \right)^2 } \right) = - \frac{P}{P_1} \left( \frac{
\int_0^{T-t} d\theta \sigma(\theta) }{ \int_0^{T_1-t} d\theta  \sigma(\theta)  }
\right)\label{eq:nhed}  
\end{eqnarray}
The popular exponential volatility function\footnote{This volatility
  function is commonly used as it lets the spot rate $r(t)$ follow a
  Markov process. See \cite{Eberlein}.} $\sigma(\theta) = \sigma
e^{-\lambda \theta}$ allows a comparison between our field theory
solutions and previous research. Under the assumption of exponential
volatility, equation (\ref{eq:nhed}) becomes
\begin{eqnarray}
\label{eq:nhedjt} 
\Delta_1 &=& - \frac{P}{P_1} \left( \frac{  1-e^{-\lambda(T-t)} }{  1-e^{-\lambda(T_1-t)}}  \right) 
\end{eqnarray}
Equation (\ref{eq:nhedjt}) coincides with the ratio of hedge
parameters found as equation 16.13 of Jarrow and Turnbull
\cite{JTtext}. In terms of their notation
\begin{eqnarray}
\Delta_1 &=& - \frac{P(t,T)}{P(t,T_1)} \left( \frac{X(t,T)}{X(t,T_1)} \right) \label{eq:jth}
\end{eqnarray}

For emphasis, the following equation holds in a one factor HJM
model\footnote{Note that this result depends on the fact that the
  spot rate $r(t)$ is Markovian and therefore only applies to either a
  constant or exponential volatility function.}
\begin{equation}
\frac{\partial \left[ P(t,T) + \Delta_1 P(t,T_1) \right]}{\partial
  r(t)} =0
\end{equation}
which is verified using equation (\ref{eq:jth}) and results found on
pages 494-495 of Jarrow and Turnbull \cite{JTtext}
\begin{eqnarray}
\frac{\partial \left[ P(t,T) + \Delta_1 P(t,T_1) \right]}{\partial r(t)} &=& 
-P(t,T) X(t,T) - \Delta_1 P(t,T_1) X(t,T_1) \nonumber \\
&=& -P(t,T) X(t,T) + P(t,T) X(t,T) = 0 \nonumber
\end{eqnarray}
When $T_1 =T$, the hedge parameter equals minus one. Economically,
this fact states that the best strategy to hedge a bond is to short a
bond of the same maturity. This trivial approach reduces the residual
variance in equation (\ref{eq:res}) to zero as $\Delta_1=-1$ and
$P=P_1$ implies $L_1=M_{11}$. Empirical results for nontrivial hedging
strategies are found in the next subsection where the calibrated
propagator is used.

\begin{figure}[h]
  \centering
  \epsfig{file=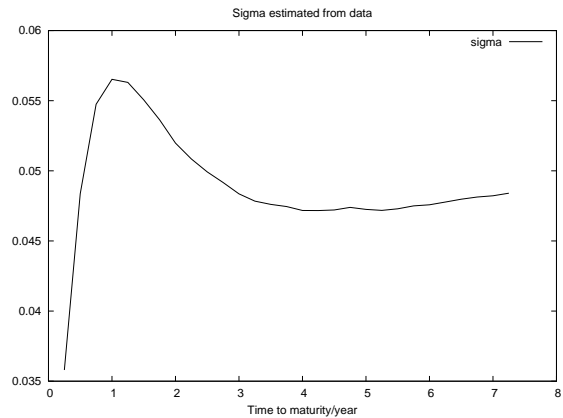, height=8cm, angle=-90}
  \caption{Implied volatility function (unnormalized) for constant
    rigidity model using market data}
  \label{fig:sigmabaaquie}
\end{figure}

\subsection{Semi-empirical results : Constant Rigidity model}

\begin{figure}[h]
  \centering
  \epsfig{file=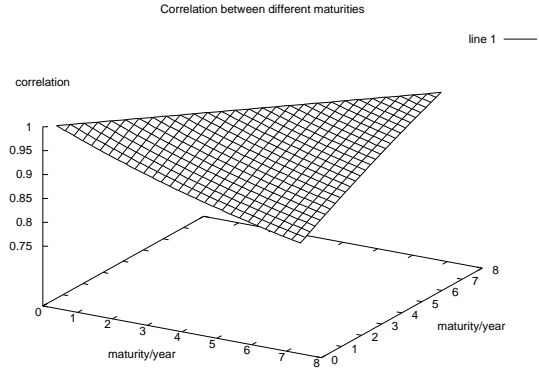, height=8cm, angle=-90}
  \caption{Propagator Implied by the constant rigidity field theory model with $\mu=0.06/\mathrm{year}$}
  \label{fig:prop}
\end{figure}

\begin{figure}[h]
  \centering
  \epsfig{file=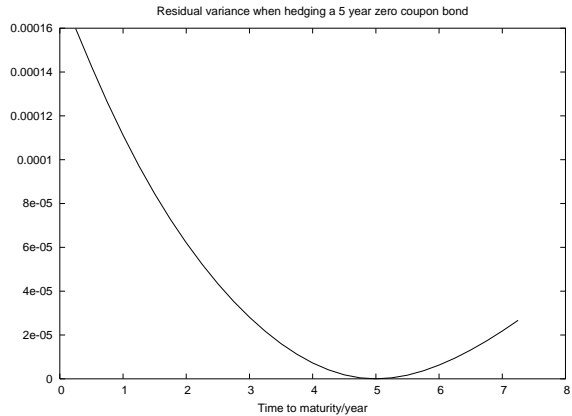, height=8cm, angle=-90}
  \caption{Residual variance for five year bond versus bond maturity used to hedge}
  \label{fig:resvariance1}
\end{figure}

\begin{figure}[h]
  \centering
  \epsfig{file=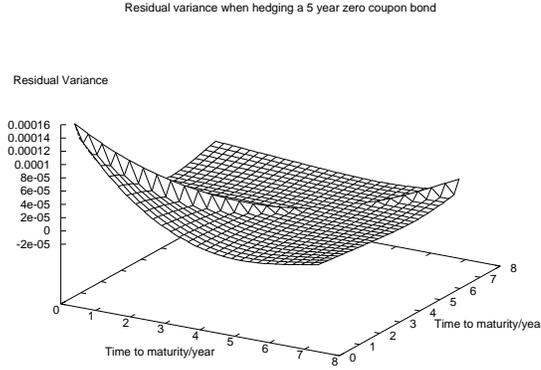, height=8cm, angle=-90}
  \caption{Residual variance for five year bond versus two bond maturities used to hedge}
  \label{fig:resvariance2}
\end{figure}

\begin{figure}[h]
  \centering
  \epsfig{file=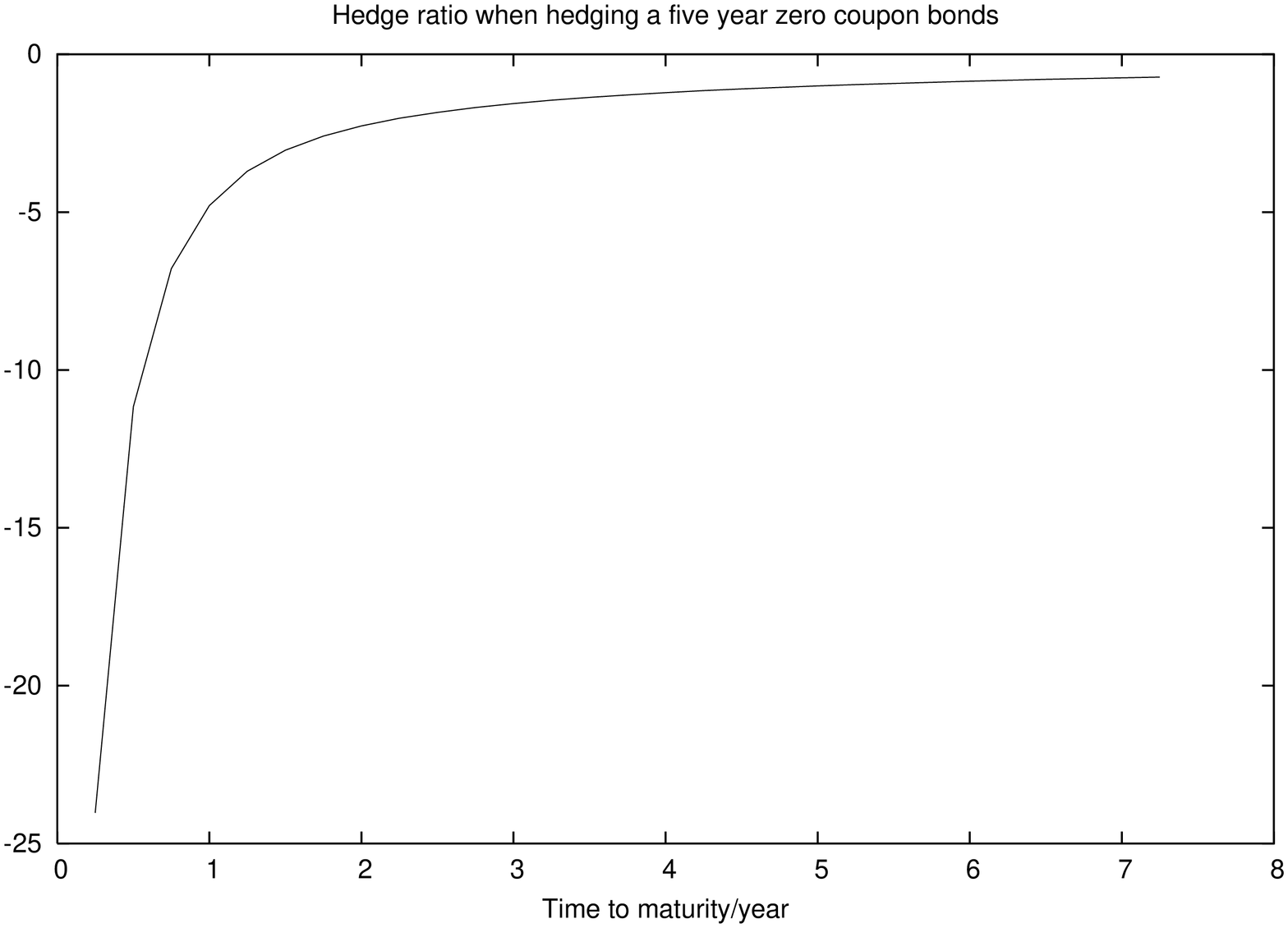, height=8cm, angle=-90}
  \caption{Hedge ratios for five year bond}
  \label{fig:hedgeratios5yr}
\end{figure}

The empirical estimation of parameters for the field theory model was
explained in detail in Baaquie and Srikant \cite{BaaqSri}. For this
subsection, we use the function $\sigma$ estimated for the constant
rigidity model from market data. The function $\sigma$ is plotted in
figure \ref{fig:sigmabaaquie}.

This approach preserves the closed form solutions for hedge parameters
and futures contracts illustrated in the previous subsection. However,
the original finite factor HJM model cannot accommodate an empirically
determined propagator since it is automatically fixed once the HJM
volatility functions are specified. Later in this subsection, we will
see how the empirical propagator modifies the results of this
subsection. The implied propagator for the empirically fitted value of
$0.06/yr$ for $\mu$ is shown in figure \ref{fig:prop}.

The reduction in variance achievable by hedging a five year bond with
other bonds is the focus of this subsection. We take the current
forward rate curve to be flat and equal to 5\% throughout. The initial
forward rate curve does not affect any of the qualitative results. The
results can also be easily extended to other bonds.  The residual
variances for one and two bond hedged portfolios are shown in figures
\ref{fig:resvariance1} and \ref{fig:resvariance2}. The calculation of
the integrals involved was done using simple trapezoidal integration
as the data is not exceptionally accurate in the first place. Secondly
and more importantly, the errors involved will largely cancel
themselves out, hence the difference in the variances is still quite
accurate. For example, in figures \ref{fig:resvariance1} and
\ref{fig:resvariance2}, we can see that in the case of perfect
hedging, we get exactly zero residual variance which shows that the
errors tend to cancel. The parabolic nature of the residual variance
is because $\mu$ is constant. A more complicated function would
produce residual variances that do not deviate monotonically as the
maturity of the underlying and the hedge portfolio increases although
the graphs appeal to our economic intuition which suggests that
correlation between forward rates decreases monotonically as the
distance between them increases as shown in figure \ref{fig:prop}.
Observe that the residual variance drops to zero when the same bond is
used to hedge itself, eliminating the original position in the
process. The corresponding hedge ratios are shown in figure
\ref{fig:hedgeratios5yr}.

It is also interesting to note that hedging by two bonds, even very
closely spaced ones, seems to be bring significant additional
benefits. This can be seen in figure \ref{fig:resvariance2} where the
diagonal $\theta=\theta'$ represents hedging by one bond. The residual
variance there is higher than the nearby points in a discontinuous
manner. 

\begin{figure}[h]
  \centering
  \epsfig{file=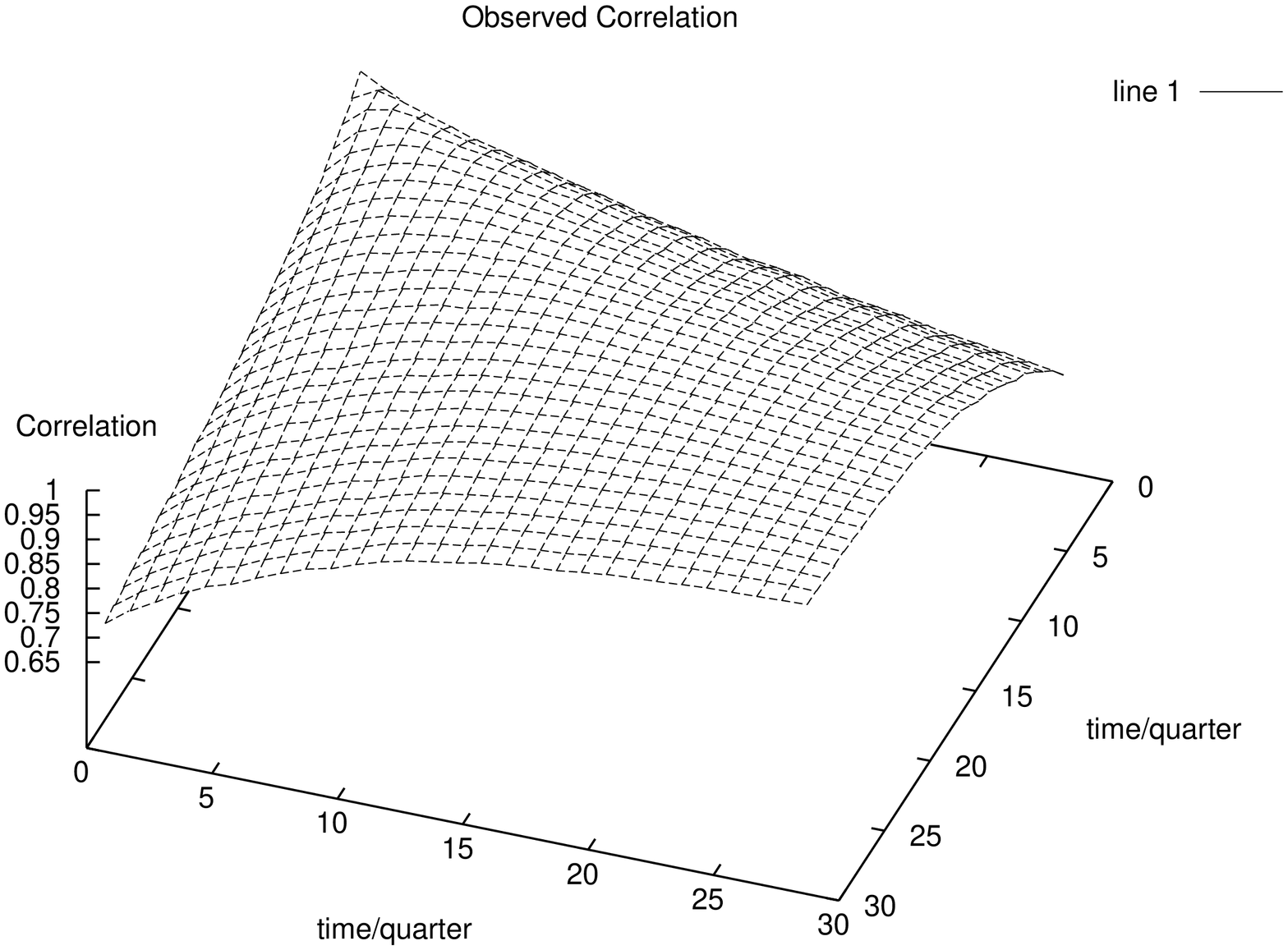, height=8cm, angle=-90}
  \caption{The propagator implied by the market data}
  \label{fig:actprop}
\end{figure}

\begin{figure}[h]
  \centering
  \epsfig{file=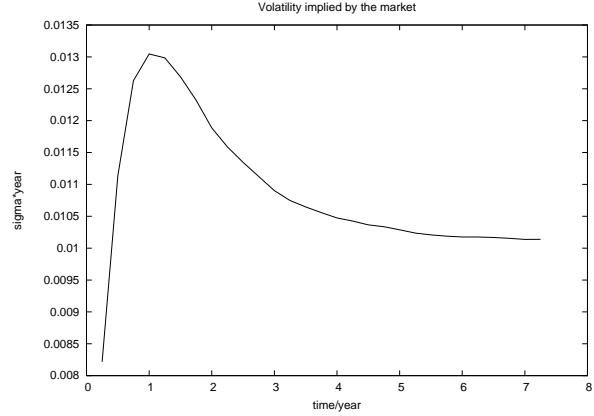, height=8cm, angle=-90}
  \caption{The volatility implied by the market data when using the empirical propagator}
  \label{fig:actprop_sigma}
\end{figure}

\begin{figure}[h]
  \centering
  \epsfig{file=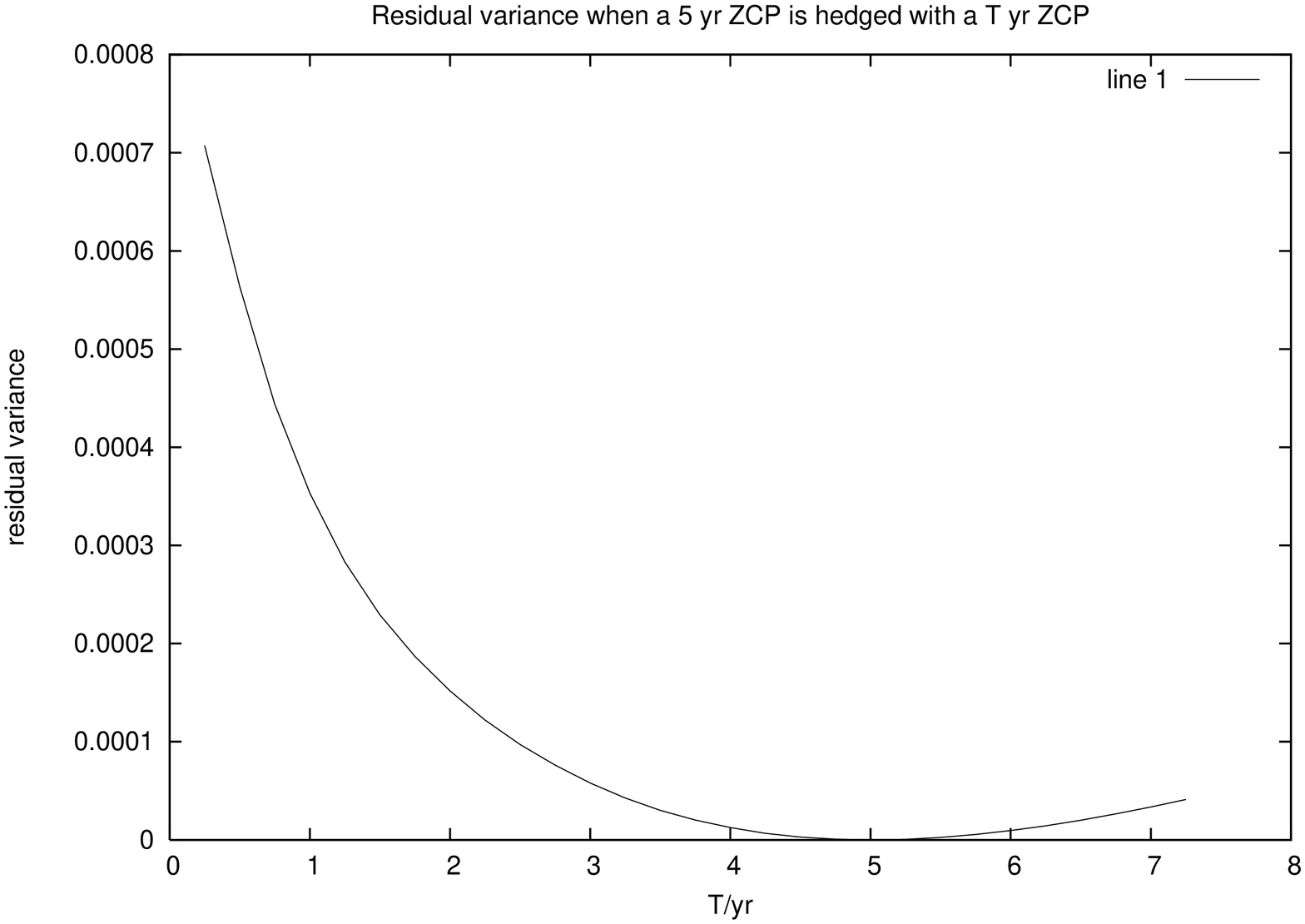, height=8cm, angle=-90}
  \caption{The residual variance when a five year bond is hedged with one bond}
  \label{fig:actprop_hedge1bond}
\end{figure}

\begin{figure}[hb]
  \centering
  \epsfig{file=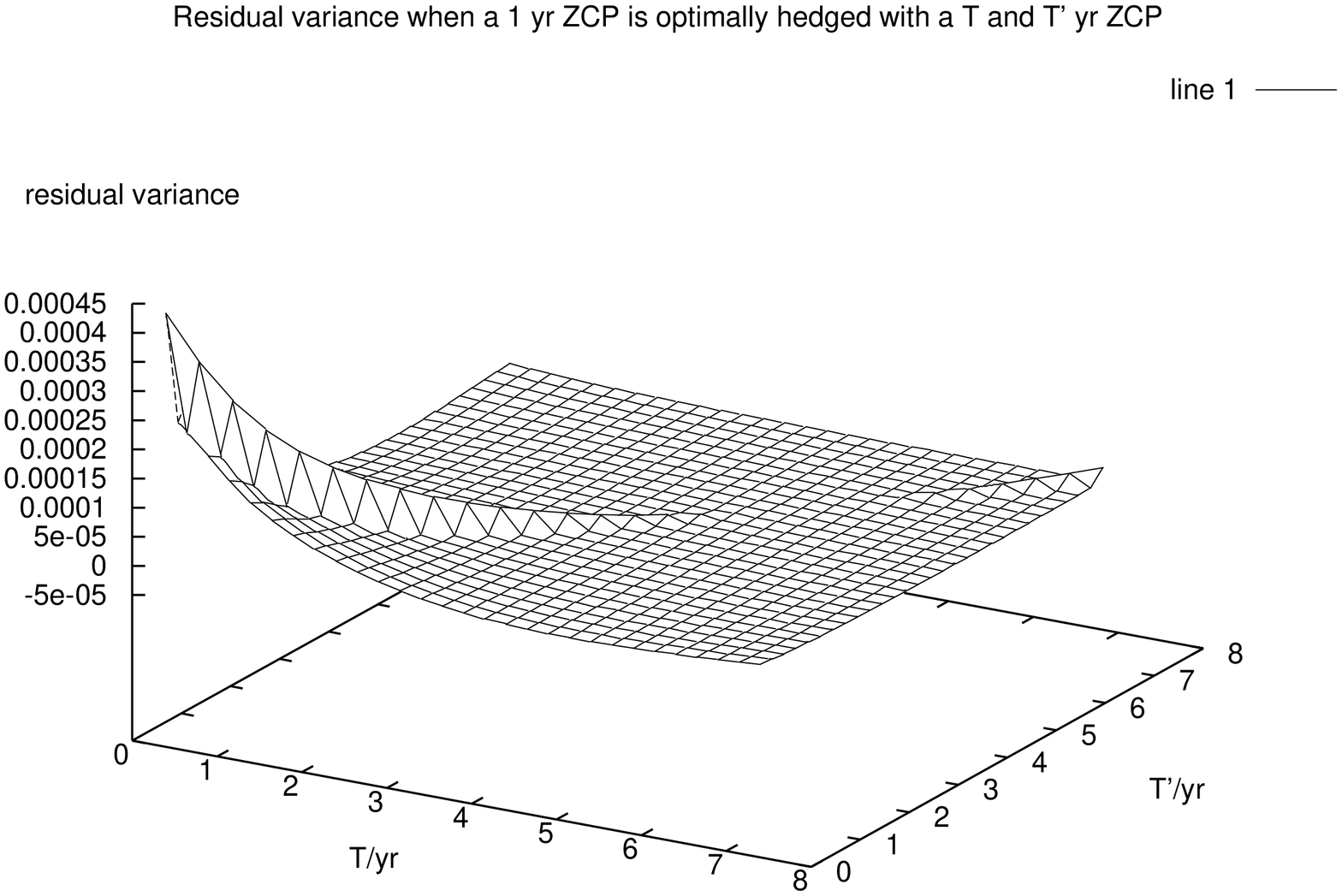, height=8cm, angle=-90}
  \caption{The residual variance when a five year bond is hedged with two bonds}
  \label{fig:actprop_hedge2bonds}
\end{figure}

\begin{figure}[h]
  \centering
  \epsfig{file=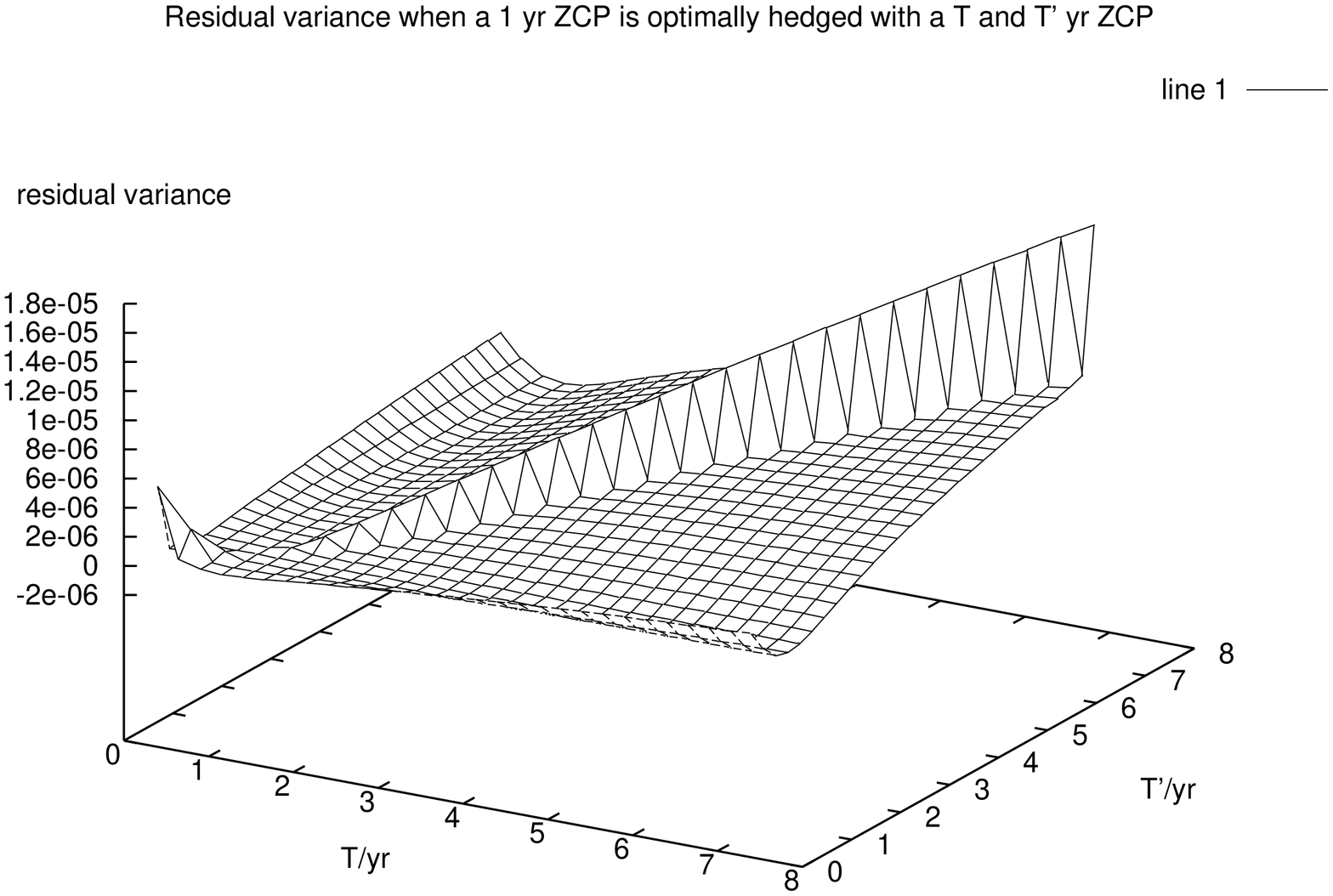, height=8cm, angle=-90}
  \caption{The residual variance when a one year bond is hedged with two bonds}
  \label{fig:actprop_hedge1yr2bonds}
\end{figure}

We now present the results for the actual propagator found from the
data which is graphed in figure \ref{fig:actprop}. The residual
variance when a five year bond is hedged with one and two bonds bond
is shown in figures \ref{fig:actprop_hedge1bond} and
\ref{fig:actprop_hedge2bonds}. We can see from figure
\ref{fig:actprop_hedge2bonds} that, when the market propagator is
used, the advantage of using more than one bond to hedge is
significantly higher. This is because of the nature of the correlation
structure in figure \ref{fig:actprop}. We see that the correlations of
innovations of nearby forward rates of higher maturity is
significantly higher in the market propagator, making hedging with
more than one bond more useful. This is even more pronounced when
hedging a short maturity bond with longer maturity ones. We can see
this from figure \ref{fig:actprop_hedge1yr2bonds} which shows the
residual variance when a one year bond is hedged with two bonds where
the calculation is done using the empirical propagator. The effect of
this higher correlation among forward rates of higher maturity can
also be seen in figure \ref{fig:actprop_hedge2bonds} where the
residual variance rises much more slowly when the hedging bonds are
chosen to be of higher maturity than the hedged bond. 

\subsection{Hedging with futures}
We can carry out an analysis very similar to subsection
\ref{theo_hedging} to find the optimal hedge ratios when hedging a
bond with futures contracts on the same or other bonds. In this case,
there is no trivial solution to the hedging problem as when bonds were
hedged with other bonds. Further, since this method of hedging is much
more practical in reality, the results will be more interesting.
Proceeding as in subsection \ref{theo_hedging}, we compute the
appropriate hedge parameters for futures contracts. The futures price
${\cal F}(t,t_*,T)$ in terms of the forward price $\frac{P(t, T)}{P(t,
  t_*)} = e^{- \int_{t_*-t}^{T-t} d\theta f(t,\theta)}$ and the \textit{deterministic}
quantity $\Omega_{\cal F}(t,t_*,T)$ which is given by \cite{Baaquie}
\begin{equation}
  \Omega_{\cal F}(t_0,t_*,T)= -\sum_{i=1}^N \int_{t_0}^{t_*}dt \int_0^{t_*-t}d\theta
  \sigma_i(t,\theta)\int_{t_*-t}^{T-t} d\theta' \sigma_i(t,\theta')
\end{equation}
The dynamics of the futures price $d{\cal F}(t,t_*,T)$ is thus given
by
\begin{equation}
  \frac{d{\cal F}(t,t_*,T)}{{\cal F}(t,t_*,T)} = d\Omega_{\cal
    F}(t,t_*,T) - \int_{t_*-t}^{T-t} d\theta df(t, \theta) 
\end{equation}
which implies
\begin{equation}
 \frac{ d{\cal F}(t,t_*,T) - E[d{\cal F}(t,t_*,T)]}{{\cal F}(t,t_*,T)}
 = -dt \int_{t_*-t}^{T-t} d\theta \sigma(\theta) A(t, \theta) 
\end{equation}
Squaring both sides as before leads to the instantaneous variance of
the futures price
\begin{equation}
  Var[d{\cal F}(t,t_*,T)] = dt {\cal F}^2(t,t_*,T) \int_{t_*-t}^{T-t} d\theta
  \int_{t_*-t}^{T-t} d\theta' \sigma (\theta) D(\theta, \theta') \sigma(\theta') 
\end{equation}

Let ${\cal F}_i$ denote the futures price ${\cal F}(t,t_*,T_i)$ of a
contract expiring at time $t_*$ on a zero coupon bond maturing at time
$T_i$. The hedged portfolio in terms of the futures contract is given
by
\begin{equation}
  \Pi(t) = P + \sum_{i=1}^N \Delta_i {\cal F}_i
\end{equation}
where ${\cal F}_i$ represent observed market prices. For notational
simplicity, define the following terms
\begin{eqnarray}
  \label{eq:M_futures}
        L_i &=& P {\cal F}_i \int_{t_*-t}^{T_i-t} d\theta \int_0^{T-t}
        d\theta' \sigma (\theta) D(\theta,\theta';T_{FR}) \sigma
        (\theta') \nonumber \\ 
  M_{ij} &=& {\cal F}_i{\cal F}_j \int_{t_*-t}^{T_i-t} d\theta \int_{t_*-t}^{T_j-t}
  d\theta' \sigma (\theta) D(\theta,\theta'; T_{FR}) \sigma (\theta') \nonumber 
\end{eqnarray}

The hedge parameters and the residual variance when futures contracts
are used as the underlying hedging instruments have identical
expressions to those in (\ref{eq:resultdelta}) and
(\ref{eq:result_var}) but are based on the new definitions of $L_i$ and
$M_{ij}$ above. Computations parallel those in section
\ref{theo_hedging}. 

To explicitly state the results, the hedge parameters for a futures
contract that expires at time $t_*$ on a zero coupon bond that matures
at time $T_i$ equals
\begin{eqnarray}
  \Delta_i &=& - \sum_{j=1}^N L_j M_{ij}^{-1} \nonumber 
\end{eqnarray}
while the variance of the hedged portfolio equals
\begin{eqnarray}
  V &=& P^2 \int_0^{T-t} d\theta \int_0^{T-t} d\theta' \sigma(\theta)
  \sigma(\theta') D(\theta,\theta';T_{FR}) - \sum_{i=1}^N \sum_{j=1}^N
  L_i M_{ij}^{-1} L_j \nonumber 
\end{eqnarray}
for $L_i$ and $M_{ij}$ as defined above.

\begin{figure}[h]
  \centering
  \epsfig{file=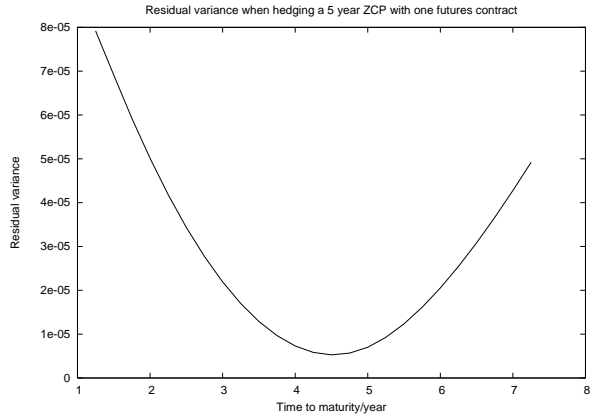, height=8cm, angle=-90}
  \caption{Residual variance for a five year bond hedged with a one year futures contract on a T maturity bond}
  \label{fig:resvar5yrfutures}
\end{figure}

\subsection{Semi-empirical results for hedging with futures}
We first present results for the propagator fitted for the constant
rigidity model as for the bonds. The initial forward rate curve is
again taken to be flat and equal to 5\%. We also fix the expiry of the
futures contracts to be at one year from the present. This is a long
enough time to clearly show the effect of the expiry time as well as
short enough to make practical sense as long term futures contracts
are illiquid and unsuitable for hedging purposes.

\begin{table}[h]
  \centering
  \begin{tabularx}{\linewidth}{|>{\setlength{\hsize}{0.5\hsize}}X|>{\setlength{\hsize}{2\hsize}}X|>{\setlength{\hsize}{0.5\hsize}}X|} \hline
    Number & Futures Contracts (Hedge Ratio) & Residual
    Variance\\ \hline \hline
    0 & none & $1.82 \times 10^{-3}$ \\ \hline
    1 & $4.5$ years ($-1.288$) & $5.29\times 10^{-6}$\\ \hline
    2 & 5 years ($-0.9347$), $1.25$ years ($-2.72497$) & $1.58 \times
    10^{-6}$\\ \hline
    3 & 5 years ($-0.95875 $), 1.5 years (1.45535), 1.25 years
    ($-5.35547$) & $1.44 \times 10^{-6}$\\ \hline
  \end{tabularx}
  \caption{Residual variance and hedge ratios for a five year
    bond hedged with one year futures contracts.}
  \label{tab:resvar_futures}
\end{table}

\begin{figure}[h]
  \centering
  \epsfig{file=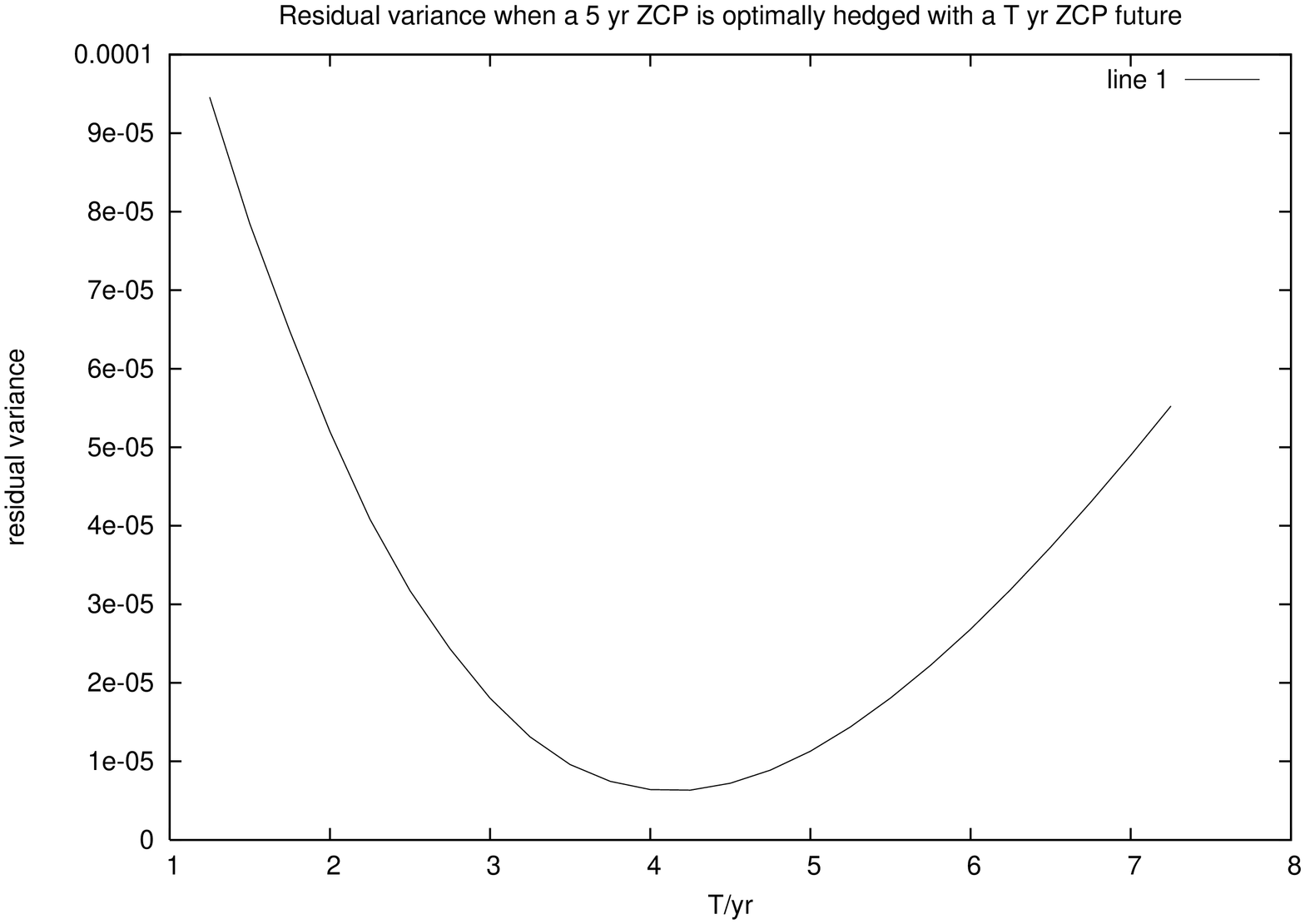, height=8cm, angle=-90}
  \caption{Residual variance when a five year bond is hedged with a one year futures contract on a T maturity bond}
  \label{fig:actprop_resvarfutures}
\end{figure}

\begin{table}[h]
  \centering
  \begin{tabularx}{\linewidth}{|>{\setlength{\hsize}{0.5\hsize}}X|>{\setlength{\hsize}{2\hsize}}X|>{\setlength{\hsize}{0.5\hsize}}X|} \hline
    Number & Futures Contracts (Hedge Ratio) & Residual
    Variance\\ \hline \hline
    0 & None & $1.74 \times 10^{-3}$\\ \hline
    1 & 4.25 years ($-0.984$) & $6.34 \times 10^{-6}$\\ \hline
    2 & 1.25 years ($-3.84577$), 5.5 years ($-0.76005$) & $2.26 \times
    10^{-6}$\\ \hline
    3  & 1.25 years ($-8.60248$), 1.5 years (2.84177), 5.25
    ($-0.85915$) & $1.95 \times 10^{-6}$ \\ \hline
  \end{tabularx}
  \caption{Residual variance and hedge ratios for a five year
    bond hedged with one year futures contracts.}
  \label{tab:actprop_resvar_futures}
\end{table}

The calculations were done using simple trapezoidal integration as
explained previously. This is sufficient for our purposes as the
fitted values for $\sigma$ and $D$ shown in figures
\ref{fig:actprop_sigma} and \ref{fig:actprop} are reasonably but not
exceptionally accurate and we are more interested in the qualitative
behaviour of the residual variance and hedge parameters.

The residual variance achieved when hedging a five year bond with one
futures contract is shown in figure \ref{fig:resvar5yrfutures}. The
optimal hedge ratios and the resulting residual variances when hedging
with two and three futures are shown in table
\ref{tab:resvar_futures}. These were obtained by systematically
tabulating all possible combinations of bonds with intervals of three
months in the maturity direction, tabulating the residual variance for
each and finding the best combination. 

Firstly, we note that the hedging is very effective even when one
futures contract is used reducing the variance by a factor of over
three hundred. Secondly, we also note that the most effective hedging
is {\em not} obtained by shorting the futures corresponding to the
same bond but one with a slightly lower maturity. This is due to the
correlation structure of the forward rates. However, when two futures
contracts are used, we see that one of the optimal contracts is the
future on the same bond as well as a very short maturity futures which
is probably due to the short end of the forward rate curve which does
influence the bond but not the futures. Since the shortest maturity
futures contract is probably likely to have the highest correlation
with this part of the forward rate curve, it seems reasonable to
select this futures contract to balance the effect on the bond from
this part of the forward rate curve. This is indeed the case as seen
in table \ref{tab:resvar_futures}. We also note that there is very
little extra improvement as we use more than two futures.

We now present the same results using the empirical propagator
directly. The residual variance when one futures contract is used for
hedging is shown in figure \ref{fig:actprop_resvarfutures}. The
optimal hedging futures, hedge ratios and residual variances are shown
in table \ref{tab:actprop_resvar_futures}. We see that for the actual
propagator, the optimal hedging futures are even farther from the
actual underlying bond when compared to the optimal values using the
fitted propagator. 

\section{Finite time hedging}
The case of finite time hedging is considerably more complicated. We
will only do the hedging of bonds with other bonds as the calculations
for minimizing variance can be done exactly. We will not do hedging of
bonds with futures even though this can also be solved exactly for
minimizing the variance as it does not add much extra insight for
finite time. To see this, consider hedging with a futures contract on
a zero coupon bond of duration $T$ that matures at the same as the
hedging horizon. This gives exactly the same result as hedging with a
bond of the same maturity $T$. Therefore, we gain nothing by carrying
out that calculation. 

The following calculation proceeds efficiently because of the use of
path integral techniques which are very useful for such problems. To
be able to optimally hedge bonds with other bonds in the sense of
having a minimal residual variance, we need to the covariance between
the final values of bonds of different maturities. To calculate this
covariance, we will first find the joint probability density function
for $N$ bonds at the hedging horizon. Let us denote the initial time
by $0$, the hedging horizon by $t_1$ and the maturities of the bonds
by $T_i$. Making use of (\ref{eq:partition}) we obtain the joint
distribution of the quantities $G_i=\int_{t_1}^{T_i} dx (f(t_1,x) -
f(0,x))$\footnote{Due to the definition of this quantity, it is easier
  to carry out the calculations for the finite case with the $x$
  rather than the $\theta$ variable, hence we use this variable in
  this section} which represent the logarithms of the ratios of final
value of the bonds to the value of their forward prices for the final
time at the initial time. In other words
\begin{equation}
  G_i = \ln \left(\frac{P(t_1, T)P(t_0, t_1)}{P(t_0, T)}\right) = \ln
  \left(\frac{P(t_1, T)}{F(t_0, t_1, T)}\right)
\end{equation}
The calculation proceeds as follows
\begin{equation}
  \begin{split}
     &\langle \prod_{j=1}^N \delta (\int_{t_1}^{T_j} dx (f(t_1,x)-f(0,x)) - G_j)
    \rangle\\
    &= \int dp_j {\cal D}A \exp\left(i\sum_{j=1}^N p_j\left(\int_0^{t_1} dt
        \int_{t_1}^{T_j} dx \alpha(t,x)
    + \int_0^{t_1} dt \int_{t_1}^{T_j} dx \sigma(t,x)
        A(t,x) - G_j\right)\right)
  \end{split}
\end{equation}
which, on applying (\ref{eq:partition}) becomes 
\begin{equation}
  \begin{split}
  \int dp_j &\exp\left(-\frac{1}{2} \sum_{j=1}^N \sum_{k=1}^N p_j
    p_k \int_0^{t_1} dt \int_{t_1}^{T_j} dx \int_{t_1}^{T_k} dx'
    \sigma(t,x) D(x-t, x'-t) \sigma(t,x')\right. \\
    &\left. + i\sum_{j=1}^N p_j \left(\int_0^{t_1} dt \int_{t_1}^{T_j} dx
      \alpha(t,x) - G_j\right)\right) 
  \end{split}
\end{equation}
Performing the Gaussian integrations, we obtain the joint probability
distribution given by 
\begin{equation}
  \label{eq:jointdistri}
  (2\pi)^{-n/2} (\det B)^{-1/2} \exp\left(-\frac{1}{2} \sum_{j=1}^N
    \sum_{k=1}^N (G_j - m_j) B_{jk}^{-1} (G_k - m_k)\right)
\end{equation}
where $B$ is the matrix whose elements $B_{ij}$ are given by
\begin{equation}
  B_{ij} = \int_0^{t_1} dt \int_{t_1}^{T_i} dx \int_{t_1}^{T_j} dx'
  \sigma(t,x) D(x-t,x'-t) \sigma(t,x') 
\end{equation}
and $m_i$ is given by 
\begin{equation}
  m_i = \int_0^{t_1} dt \int_{t_1}^{T_i} dx \alpha(t,x)
\end{equation}
Hence, the quantities $G_i$ follow a multivariate Gaussian
distribution with covariance matrix $B_{ij}$ and mean $m_i$.

Having found the joint distribution of $G_i$, we can find the
covariance of the final bond prices by tabulating the expectations of
each of the bonds and the expectation of their products. The final
bond price is given by $P(t_1, T_i) = F(0, t_1, T_i) e^{G_i}$ in terms
of $G_i$. Hence, the expectation of this quantity is given by 
\begin{equation}
  \label{eq:expectbond}
  (2\pi)^{-N/2} (\det B)^{-1/2} \int dG_i F(0,t_1, T_i) e^{G_i}
  \exp\left(-\frac{1}{2} (G-m)^T B_{ij}^{-1} (G-m)\right)
\end{equation}
which gives $\cali{F}(0, t_1, T_i)$ as it must since the expectation
of the future bond price is the futures price. The expectation of the
products of the prices of two bonds $\langle P(t_1, T_i) P(t_1,
T_j)\rangle$ is given by
\begin{equation}
  (2\pi)^{-N/2} (\det B)^{-1/2} \int dG_i dG_j F(0, t_1, T_i) F(0,t_1,
  T_j) e^{G_i + G_j} \exp\left(-\frac{1}{2}H^T B^{-1} H\right)
\end{equation}
where $H$ stands for the vector $G-m$. On evaluation, this gives the
result 
\begin{equation}
  \cali{F}(0, t_1, T_i) \cali{F}(0, t_1, T_j) \exp\left(\int_0^{t_1}
    dt \int_{t_1}^{T_i} dx \int_{t_1}^{T_j} dx' \sigma(t,x) D(x-t,
    x'-t) \sigma(t,x') \right)
\end{equation}

We now consider the behaviour of the portfolio 
\begin{equation}
  P(t, T) + \sum_{i=1}^N \Delta_i P(t, T_i) 
\end{equation}
The covariance between the prices $P(t_1, T_i)$ and $P(t_i, T_j)$ is
given by
\begin{equation}
  \label{eq:Mfinitetime}
  \begin{split}
    M_{ij} =& \cali{F}(0, t_1, T_i) \cali{F}(0, t_1, T_j)\\
    &\left(\exp\left (\int_0^{t_1} dt \int_{t_1}^{T_i} dx
        \int_{t_1}^{T_j} dx' \sigma(t,x) D(x-t, x'-t)
        \sigma(t,x')\right) -1\right)
    \end{split}
\end{equation}
and the covariance between the hedged bond of maturity $T$ and the
hedged bonds is given by
\begin{equation}
  \label{eq:Lfinitetime}
  \begin{split}
  L_i =& \cali{F}(0, t_1, T) \cali{F}(0, t_1, T_i)\\
    &\left(\exp\left (\int_0^{t_1} dt \int_{t_1}^{T} dx
        \int_{t_1}^{T_i} dx' \sigma(t,x) D(x-t, x'-t)
        \sigma(t,x')\right) -1\right)
    \end{split}
\end{equation}
and the minimization of the residual variance of the hedged portfolio
proceeds exactly as in the first section. The hedge ratios are found
to be given by 
\begin{equation}
  \Delta = L^T M^{-1} 
\end{equation}
and the minimized variance is again 
\begin{equation}
  \mathrm{Var}[P(t, T)] - L^T M^{-1} L
\end{equation}
It is not too difficult to see that both $M$ and $L$ reduce to the
results in the first section if $t_1 \rightarrow 0$ (with the
covariances being scaled by $t_1$, of course). 

One very interesting difference between the instantaneous hedging and
finite time hedging is that the result depends on the value of
$\alpha$. In the calculation above, we used the risk-neutral $\alpha$
obtained for the money market numeraire. However, the market does not
follow the risk-neutral measure and it would be better to use a value
for $\alpha$ estimated for the market for any practical use of this
method. This difference is expected since in the very short term only
the stochastic term dominates making the drift inconsequential. This,
of course, is not the case for the finite time case where the drift
becomes important (it is not difficult to see that the importance of
the drift grows with the time horizon). 

\begin{figure}[h]
  \centering
  \epsfig{file=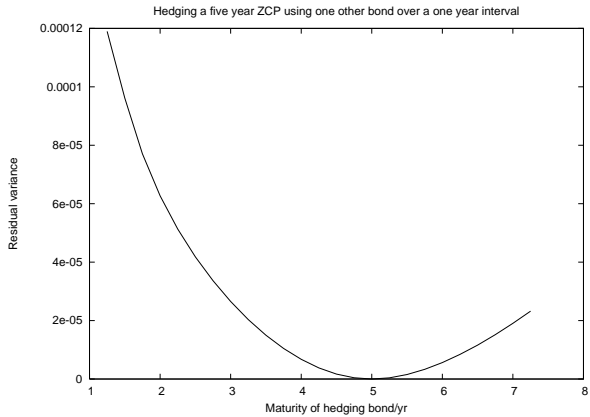, height=8cm, angle=-90}
  \caption{Residual variance when a five year bond is hedged with one
    other bond (best fit of the constant rigidity field theory model)
    with a time horizon of one year}
  \label{fig:ftime_resvar}
\end{figure}

\begin{figure}[h]
  \centering
  \epsfig{file=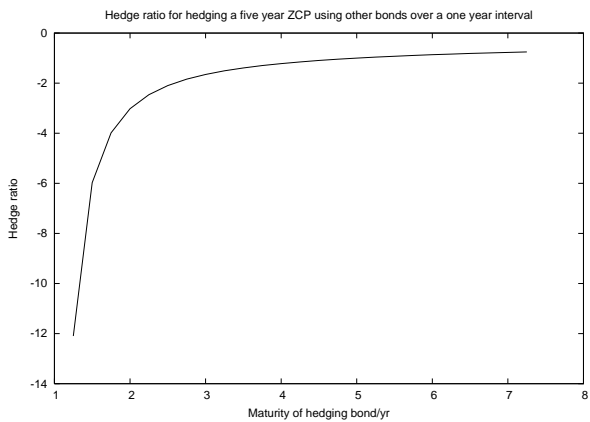, height=8cm, angle=-90}
  \caption{Hedge ratio when a five year bond is hedged with one other
    bond (best fit of the constant rigidity field theory model) with a
    time horizon of one year }
  \label{fig:ftime_hratio}
\end{figure}

\begin{figure}[h]
  \centering
  \epsfig{file=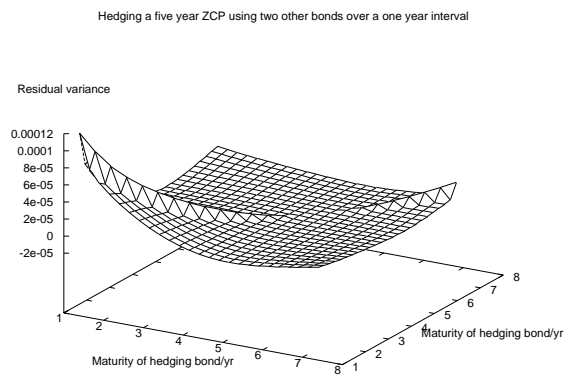, height=8cm, angle=-90}
  \caption{Residual variance when a five year bond is hedged with two
    other bonds (best fit of the constant rigidity field theory model)
    with a time horizon of one year}
  \label{fig:ftime_resvar2d}
\end{figure}

\section{Semi-empirical Results for Finite Time Hedging}
We now present the empirical results for hedging of a bond with other
bonds for both the best fit for the constant rigidity field theory
model and the fully empirical propagator. The calculation of $L$ and
$M$ were again carried out using simple trapezoidal integration and
$\sigma$ was assumed to be purely a function of $\theta = x-t$ so that
all the integrals over $x$ were replaced by integrals over $\theta$.
The bond to be hedged was chosen to be the five year zero coupon bond
and the time horizon $t_1$ was chosen to be one year.

The results for the best fit of the constant rigidity field theory
model (see figures \ref{fig:sigmabaaquie} and \ref{fig:prop}) for the
residual variance and hedge ratio for hedging with one bond are shown
in figures \ref{fig:ftime_resvar} and \ref{fig:ftime_hratio}. The
residual variance for hedging with two bonds is shown in figure
\ref{fig:ftime_resvar2d}.

\begin{figure}[h]
  \centering
  \epsfig{file=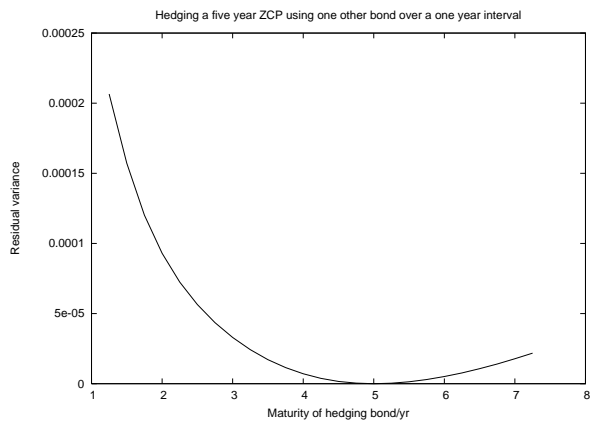, height=8cm, angle=-90}
  \caption{Residual variance when a five year bond is hedged with one
    other bond (best empirical fit) with a time horizon of one year}
  \label{fig:actprop_ftime_resvar}
\end{figure}

\begin{figure}[h]
  \centering
  \epsfig{file=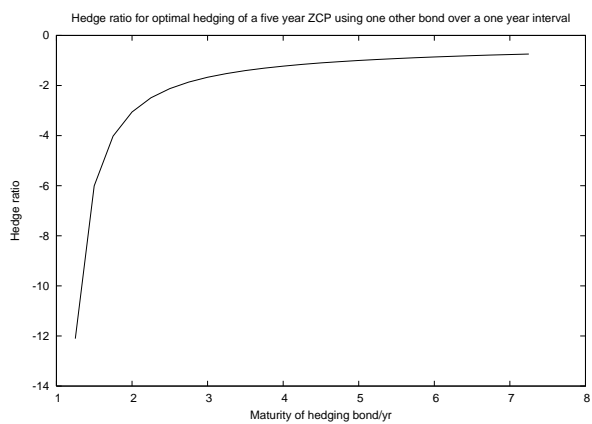, height=8cm, angle=-90}
  \caption{Hedge ratio when a five year bond is hedged with one other
    bond (best empirical fit) with a time horizon of one year }
  \label{fig:actprop_ftime_hratio}
\end{figure}

\begin{figure}[h]
  \centering
  \epsfig{file=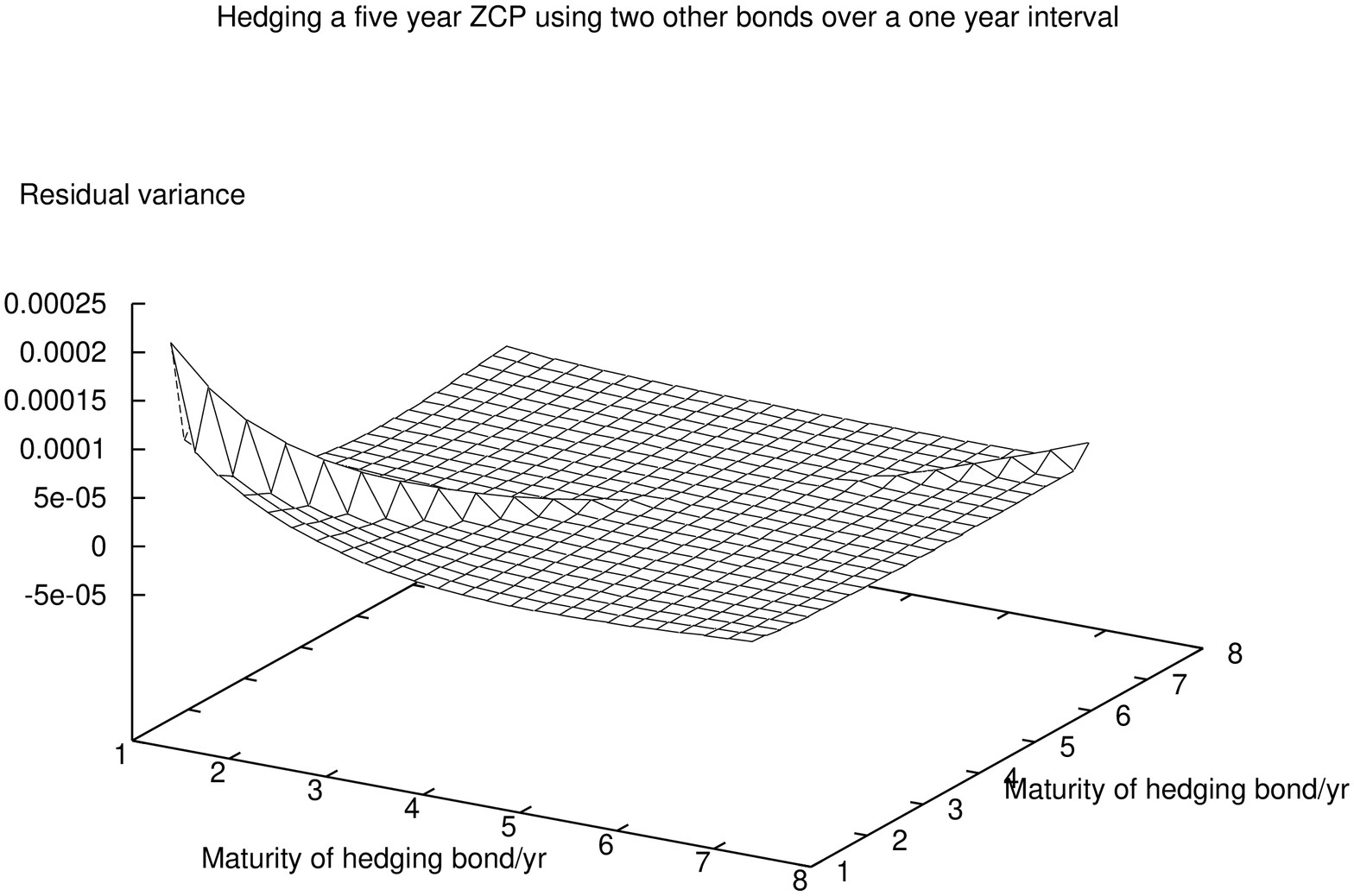, height=8cm, angle=-90}
  \caption{Residual variance when a five year bond is hedged with two
    other bonds (best empirical fit) with a time horizon of one year}
  \label{fig:actprop_ftime_resvar2d}
\end{figure}

The results for the fully empirical quadratic fit (see figures
\ref{fig:actprop} and \ref{fig:actprop_sigma}) for the residual
variance and hedge ratio for hedging with one bond are shown in
figures \ref{fig:actprop_ftime_resvar} and
\ref{fig:actprop_ftime_hratio}. The residual variance for hedging with
two bonds is shown in figure \ref{fig:actprop_ftime_resvar2d}.

One interesting result is that the actual residual variance after
hedging over a finite time horizon is lesser than naively
extrapolating from the infinitesimal hedging result. This seems to be
due to the shrinking nature of the domain as the contribution to the
variance of the bonds reduces as the time horizon increases. This is
very clear if the maturity of the bond is close to the hedging horizon
as the volatility of bonds reduces quickly as the time to maturity
approaches. Apart from this reduction, the results look very similar
to the infinitesimal case. This is probably due to the fact that the
volatility is quite small so the nonlinear effects in the covariance
matrix (\ref{eq:Mfinitetime}) are not apparent. If very long time
horizons (ten years or more) and long term bonds are considered, the
results will probably be quite different. We see by comparing figures
\ref{fig:ftime_resvar2d} and \ref{fig:actprop_ftime_resvar2d} that the
better improvement in using more than one bond to hedge when the
empirical rather than the field theory model propagator is used is
seen to be true in the finite time case as well.

\section{Conclusion}
We have shown that the field theory model offers techniques to
calculate hedge parameters for fixed income derivatives and provides a
framework to answer questions concerning the number and maturity of
bonds to include in a hedge portfolio. We have also seen how the field
theory model can be used to estimate hedge parameters for finite time
as well which is useful in practice. We have used the field theory
model calibrated to market data to show that a low dimensional basis
provides a reasonably good approximation within the framework of this
model. This shows that field theory models address the theoretical
dilemmas of finite factor term structure models and offer a practical
alternative to finite factor models. 

\section{Acknowledgements}
We would like to thank Prof. Warachka for many insightful and
illuminating discussions. We would also like to thank Jean
Phillipe-Bouchaud and Science and Finance for kindly providing us with
the data for the semi-empirical section of the study.

\bibliography{list-int}
\bibliographystyle{phaip}
\end{document}